Supporting Information

# Intramolecular Torque, an Indicator of the Internal Rotation Direction of Rotor Molecules and Similar Systems

Rui-Qin Zhang[1*], Yan-Ling Zhao[2*], Fei Qi[2], Klaus Hermann[3], and Michel A. Van Hove[2]

[1]*Department of Physics and Materials Science, City University of Hong Kong, Hong Kong SAR, China*

[2]*Institute of Computational and Theoretical Studies & Department of Physics, Hong Kong Baptist University, Hong Kong SAR, China*

[3]*Inorganic Chemistry Department, Fritz-Haber-Institut der Max-Planck-Gesellschaft, Faradayweg 4-6, 14195 Berlin, Germany*

* Authors who contributed equally to this work.

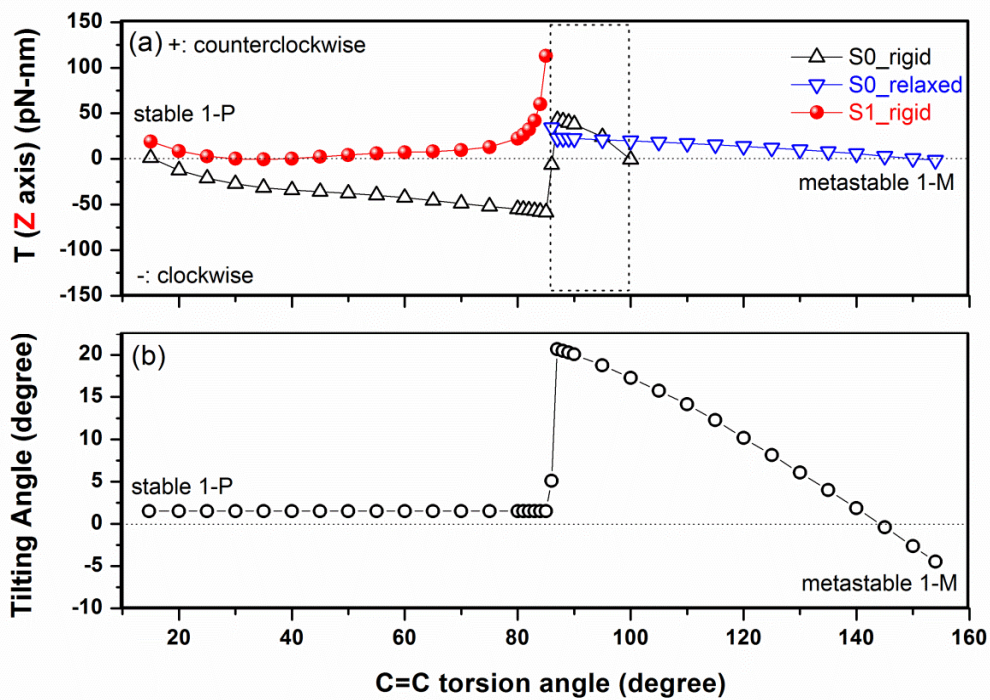

Figure S1 (a) Profiles of the S0 and S1 total torques $T$ (in units of piconewton × nanometer = $10^{-21}$ Nm) of the rotor portion of motor 1 projected onto the Z direction and (b) tilting angle of the C=C axis relative to Z direction as a function of the twisting mode ($\theta$) from 1-P to 1-M, assuming rigid rotation of the rotator until 85°, then optimized rotation beyond 85° (see text for details).

Table S1. Comparisons of the excitation energies in eV and corresponding wavelengths in nm (in parentheses) of the bistable isomers 1-P and 1-M at various levels of calculations

| Method and basis set | 1-P | 1-M |
|---|---|---|
| TD-CAM-B3LYP/6-31+G* | 3.77 (329) | 3.40 (364) |
| TD-BH&HLYP/6-31G*[28] | 3.78 (328) | 3.39 (366) |
| SA-REBH&HLYP/6-31G*[28] | 3.80 (327) | 3.40 (365) |
| OM2/GUGA-MRCI[28] | 3.93 (315) | 3.75 (331) |

# Intramolecular Torque, an Indicator of the Internal Rotation Direction of Rotor Molecules and Similar Systems

Rui-Qin Zhang[1*], Yan-Ling Zhao[2*], Fei Qi[2], Klaus Hermann[3], and Michel A. Van Hove[2]

[1]*Department of Physics and Materials Science, City University of Hong Kong, Hong Kong SAR, China*
[2]*Institute of Computational and Theoretical Studies & Department of Physics, Hong Kong Baptist University, Hong Kong SAR, China*
[3]*Inorganic Chemistry Department, Fritz-Haber-Institut der Max-Planck-Gesellschaft, Faradayweg 4-6, 14195 Berlin, Germany*

**Abstract**
Torque is ubiquitous in many molecular systems, including collisions, chemical reactions, vibrations, electronic excitations and especially rotor molecules. We present a straightforward theoretical method based on forces acting on atoms and obtained from atomistic quantum mechanics calculations, to quickly and qualitatively determine whether a molecule or sub-unit thereof has a tendency to rotation and, if so, around which axis and in which sense: clockwise or counterclockwise. The method also indicates which atoms, if any, are predominant in causing the rotation. Our computational approach can in general efficiently provide insights into the rotational ability of many molecules and help to theoretically screen or modify them in advance of experiments or before analyzing their rotational behavior in more detail with more extensive computations guided by the results from the torque approach. As an example, we demonstrate the effectiveness of the approach using a specific light-driven molecular rotary motor which was successfully synthesized and analyzed in prior experiments and simulations.

**Introduction**
The concept of torque [1-5] – the induction of rotation, as opposed to translation, by external or internal forces – is ubiquitous in dynamic molecular phenomena such as collisions, chemical reactions, vibrations and electronic excitations. The concept is especially useful for rotor molecules, in which energy from some source (e.g. chemical reaction, electron-electron or optical excitation) is converted to rotational motion [6-11] through the action of a net torque on the whole molecule or on a part of the molecule (often called rotator), as in the bacterial flagella, for example.

However, a molecular rotator is rarely rigid: it will be flexible to some degree, changing its shape as well as its electronic structure while it rotates. Thus, rigid-body mechanics do not apply directly: the classical moment of inertia matrix and the principal axes of a free molecule or rotator can be defined, but they are physically meaningful only at the start of a rotation and change continuously as the molecule changes its shape and state. (This situation is analogous to a spacecraft [12-13] in which a multitude of components can rotate irregularly, e.g. antennae, solar panels, astronauts, liquids, gases, pumps and fans, affecting the spacecraft's orientation and spin, and thus complicating navigation.)

We shall here utilize the concept of torque in rotor molecules to predict whether a particular molecule prepared in a specific state (e.g. photoexcited or twisted in a collision) will internally rotate or not. Also, the torque vector will make it possible to predict the dominant direction of rotation: around which axis and in what sense: clockwise vs. counterclockwise. As an illustration, we take a well-studied rotor molecule (called 9-(2,4,7-trimethyl-2,3-dihydro-1H-inden-1-ylidene)-9H-fluorene (motor 1 in this work)) [14-16] consisting of a rotator and a stator and ask whether and in which direction the rotator will turn after photoexcitation.

We are asking for a qualitative answer, not e.g. a precise speed of rotation about a precisely determined axis. The aim is to qualitatively sketch the rotational evolution of the system: will it rotate clockwise or counterclockwise or have negligible rotation, and what will be the approximate direction of the axis of rotation? It is possible to obtain qualitative answers by a single total-energy calculation in the molecule's initial state (which yields forces on all the atoms and, hence, atom-derived torques after choosing a suitable rotation center (pivot point)): that qualitative information may, for example, be sufficient to screen out unpromising candidate molecules from a search for useful rotor molecules, saving the time and expense of more detailed simulations of these candidates; likewise, this torque method allows quickly exploring alternate molecules.

Furthermore, if desired, one may proceed to explore the subsequent rotation with a few similar qualitative calculations after rotation has been initiated.

To study either initial rotation or subsequent rotation, the torque (referred to a selected pivot point) can be very easily calculated from the forces acting on the individual atoms, especially if an obvious axis of rotation is available, such as a single strong atom-atom bond acting as an "axle" between stator and rotor. Even when the preferred axis of rotation is not so obvious, considering the torque components along three orthogonal axes will suffice to indicate the general direction of initial rotation, or the absence of a tendency to rotate.

Using the torque can result in very large savings in computer time, both in the initial configuration of the molecule(s) and during subsequent rotation. As indicated above, in the initial state, a single total-energy calculation will show whether the molecule is prone to rotation or not, and, if so, in which approximate direction. The traditional approach requires multiple total-energy calculations to obtain this information. The prediction of the rotation direction by the torque method enables the detailed exploration to be limited to a much smaller part of the multi-dimensional configuration space, avoiding a blind search in unpromising regions.

The torque calculation can also be applied to more complex systems than simple stator+rotator molecules: for example, a molecule may have multiple rotators (including a rotator within a rotator), or the molecule may have a more complex link between stator and rotator than a single bond (e.g. three consecutive CC bonds in stilbene). Then the same calculation can yield torques around each bond or rotation axis to evaluate its individual tendency and direction of rotation, at negligible extra computational cost since all the atomic positions and forces are available for each atom from a single calculation.

The torque approach can also easily be applied to molecular collisions, chemical reactions, vibrations or electronic excitations: for given atomic positions (e.g. distortion of a molecule by a collision, by reaction with another molecule, by vibration or by excitation), a single total-energy calculation can obtain the forces acting on each atom and therefore lead to the torque acting on any part of the molecule(s), thus predicting the presence and direction of a resulting rotation (of course, with multiple molecules one should also consider the angular momenta of the incoming and outgoing molecules or fragments, as in standard collision dynamics).

**Methodology**

In a general approach we need to distinguish between rotational and translational motion induced by forces on atoms or groups of atoms. Translational motion is more relevant to vibrational analysis, while rotational motion is of special interest when examining the rotator properties of rotor molecules. For the present purpose, we neglect translational motion and focus on rotation.

The total torque on a rotator (or on any molecule or subunit of a molecule) is obtained as the vector $\vec{T} = \sum_i \vec{r}_i \times \vec{F}_i$, where $i$ runs over all atoms of the rotator (or other subunit of interest), $\vec{F}_i$ is the total force acting on atom $i$ due to all other atoms in the molecule (e.g. as a Pulay force obtained from a DFT calculation), and $\vec{r}_i$ is the position of atom $i$ relative to a pivot O on the chosen axis of rotation. At this stage the orientation of the axis of rotation is undecided: the actual axis of rotation is chosen by the system and will result partly from the torque, but will also be affected by the constraints imposed by any rigidity of the linkage to a stator and by the different masses of the atoms, i.e. inertia. Some degree of insight is needed to choose the pivot O. For colliding molecules or reaction fragments in vacuum, the pivot could be located at the center of mass of each separate molecule or fragment, since a free object will rotate around its center of mass. In this case, as the stator part is relatively fixed (like its name suggests), we place the pivot O on its one C atom and set the *X, Y, Z* coordinate directions (see Figure 1). In the stable conformer of motor 1, the C=C bond orientation is basically parallel with the Z direction. This setting will help us observe the deviation of the C=C axis from the Z axis during the rotation.

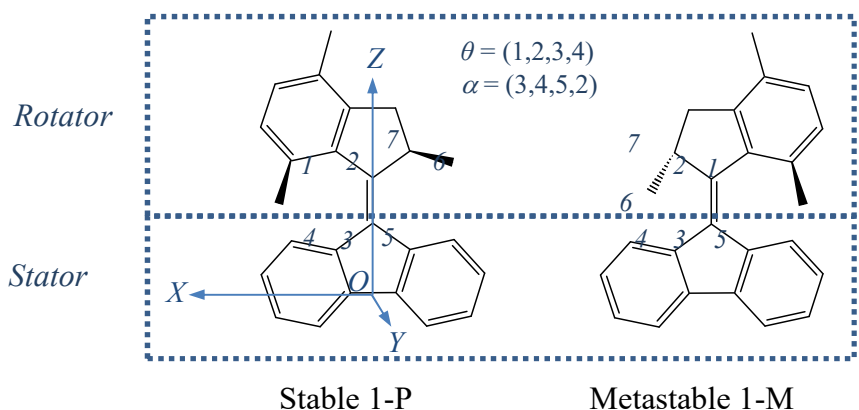

Figure 1. A schematic representation of the stable and metastable states, with labelling of key atoms and definition of two dihedral angles, of the "motor 1" molecule [14-16].

For convenience and clarity, we shall in the following discussion assume that the stator has a fixed orientation, but that it can still deform somewhat during rotation of the rotator: this may be achieved strictly by fixing the mutual orientation of a particularly important triplet of atoms of the stator. (In a free stator+rotator molecule, both stator and rotator will actually turn in opposite directions to conserve total angular momentum, so the designation of stator vs. rotator becomes arbitrary. If the stator is relatively large and bulky, or is itself rigidly attached to a larger mass, it will be nearly static.) We also assume zero temperature, i.e. no thermal vibrations.

If the total torque vector $\vec{T}$ on a rotator is small (e.g. comparable to or smaller than the average individual atomic torque), then there is only a weak tendency to rotation: this indicates that the atomic torques largely cancel each other out or are individually small. This outcome shows that the molecule actually does not behave like a rotor in its assumed configuration and state: it may not be a suitable "rotor molecule", so one may consider alternate molecules instead.

Now consider the more interesting case where the total torque vector $\vec{T}$ is larger (e.g. larger than individual atomic torques or dominated by a few atomic torques). If $\vec{T}$ is found to be parallel to the bond between stator and rotator, it is clear that the rotator will (at least initially) rotate around that bond axis, like a top spinning with a steady vertical axis on a table. The direction of rotation (clockwise or counterclockwise as seen from rotator to stator in this paper) will simply be given by the direction in which $\vec{T}$ points ("up" vs. "down"). But if $\vec{T}$ is tilted away from that bond axis, there is also a torque component that will make the rotator tilt to the side, like a spinning top tilting sideways from the vertical on a table; the rotator may then process like a tilted top, but that motion will depend also on any other deformations taking place in the molecule during the subsequent rotation.

If we wish to inspect any tilt of the rotation axis away from the bond axis (resulting in a constrained rotation about the fixed axis), we may simply take the projection $\vec{T}_\parallel$ ($\vec{T}_Z$) of the total torque $\vec{T}$ onto the $Z$ axis and use it to predict the sense of rotation about that fixed direction. Or otherwise, our torque method provides the component of the total torque perpendicular to the bond axis, $\vec{T}_\perp = \vec{T} - \vec{T}_\parallel = \vec{T}_X + \vec{T}_Y$: that perpendicular component shows the approximate direction in which this tilting would happen, and roughly how strong the tendency of such tilting is.

**Application**

We shall here illustrate the usefulness of the concept of torque for the case of a complex synthetic stator+rotator molecule (motor 1) that has already been extensively studied experimentally [14] and theoretically [15-16]. Its complexity is due to the desire to give it unidirectional motor behavior, obtained by appropriate chirality, rotational barriers and photoexcitation. The previous theoretical studies [15-16] used extensive mapping of the detailed geometry of the molecule during the rotation in order to determine its configurational trajectory in multidimensional space. We show below that we can obtain the rotational character more easily by calculating the torque at a few selected geometries. Such a quick study of the molecule's global behavior would very much reduce the need to explore large volumes of configuration space and thus would guide a subsequent detailed study more efficiently toward a full understanding of the mechanism of the system.

The molecular "motor 1" has two photo-equilibria with stable (1-P) and metastable (1-M) states, with the electronic energy difference 4.5 kcal/mol [15-16,21]. As shown in Figure 1, the upper half of motor 1 is the rotor and the lower half is the stator, which are connected together via a central C=C double bond axis. The key atoms and dihedral angles are labelled and defined in Figure 1. The fluorene-based stator has axial symmetry when it is disconnected from the rotor. Such a design splits the 360° rotation into two identical 180° rotations with a total of four steps: two identical photoisomerizations and two identical thermal relaxations, shown as a cycle in Figure 2. It can be seen that the geometries of stable 1-P and stable 1-P' are basically the same and so are those of metastable 1-M and metastable 1-M'.

In this work, we choose the CAM-B3LYP [17-19] functional with a 6-31+G* basis set to carry out the calculations conducted using the Gaussian 09 code [20]. The reliability of this functional is validated by comparing the excitation energies with reported higher levels of TD-DFT (see Table S1).

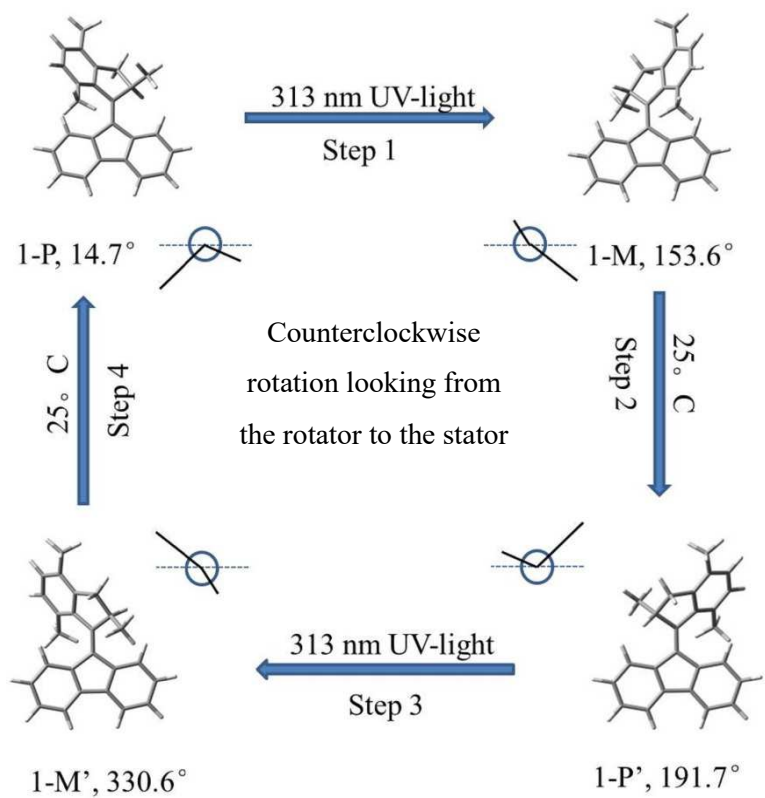

Figure 2. The two photochemical and two thermal steps of a complete 360° rotary cycle of motor 1 [14-16]. The dihedral angle $\theta$ (as a function of torque profiles in Figure 3) is displayed to track the C2=C3 twisting mode.

The rotational trajectory of "motor 1" is analyzed in refs. 15 and 16 in terms of potential energy surfaces (PES) for the ground state S0 and the excited state S1, calculated and displayed as a function of two important angles: the torsion angle $\theta$ around the linking C=C bond and a "pyramidalization angle" $\alpha$ which essentially describes the tilting of the C=C bond, as shown in Figure 1. The torsion angle $\theta$ is the main interest for this rotor molecule and will thus be our focus here.

In the initial lowest-energy stable state (1-P), there are no net forces on any atom (at T=0), so the total torque on the rotator is zero. This is true regardless of the location of the pivot O, but for the following we place the pivot at one C atom of the stator portion. The photoexcitation to the excited state S1 first leads to the Franck-Condon state in which the atomic geometry has not changed while the electronic level occupation has changed: this causes non-zero net forces on most atoms and results in a significant positive total torque on the rotator, as projected onto the C=C bond axis (see Figure 3b near the ground state's torsion angle of 15°). The positive sign implies counterclockwise rotation, looking from the rotator toward the stator. This counterclockwise rotation is induced from the decreasing slope of the PES surface of the S1 state (also seen as a decreasing slope in Figure 3a), driving the system in the counterclockwise direction.

We see here the contrast in the two approaches for obtaining the direction of rotation. The PES approach requires multiple total-energy calculations in high-dimensional search space to find a path of descending energy; it is not sufficient to observe that a particular dihedral angle between four atoms will turn counterclockwise, since the rest of the rotator could be turning in any other direction, e.g. clockwise. By contrast, the total torque uses a single total-energy calculation at the initial geometry to indicate overall rotation direction of the rotator.

For further rotation away from 1-P, the torque method can efficiently simulate a simple rigid rotation of the rotator (until steric hindrance stops it) to explore qualitatively whether there is continued tendency to rotate clockwise. At this simplest level, we obtain the total torque (and total energy) shown in Figure 3 until a torsion angle of ~100° (where weak steric hindrance starts). We see that the total torque on the rotator remains positive or close to zero, so the rotation can continue counterclockwise (the more exact PES analysis of Refs. 15 and 16 confirms a continuous downward slope in the counterclockwise direction). We have calculated the FC state every 5° or 1° for Figure 3 (with relatively rapid electronic optimization but without time-consuming geometric optimization): we could have skipped most of those points by using steps of 30°, for example, to save even more computation time.

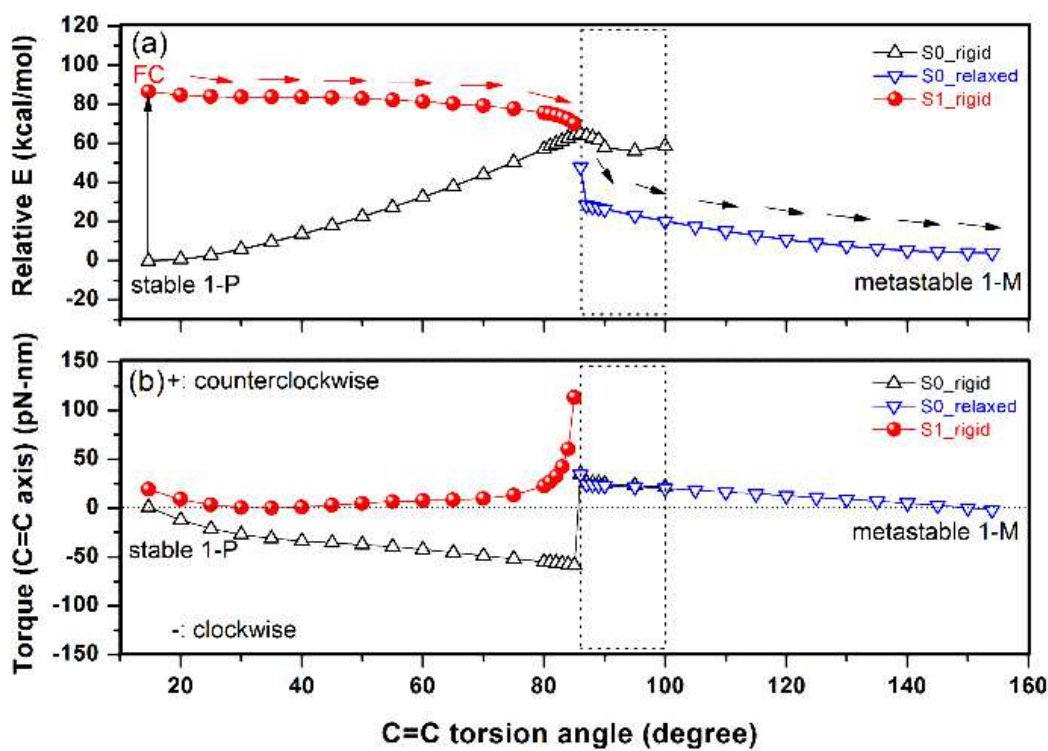

Figure 3. Profiles of the S0 and S1 PESs $E$ (a) (in kcal/mol) and total torques $T$ (in units of piconewton × nanometer = $10^{-21}$ N•m) of the rotor portion (b) of motor 1 projected onto the C=C double bond twisting mode ($\theta$) from 1-P to 1-M, assuming rigid rotation of the rotator until 85°, then optimized rotation beyond 85° (see text for details). The arrows show the expected counterclockwise system evolution.

The saddle point for twisting the C=C double bond on the ground state PES, as seen in Figure 3a, occurs near a torsion angle of 85°, causing high energies in that region. Also, near 85° the ground state S0 and the excited state S1 approach each other closely in energy (Figure 3a): this suggests the proximity of a conical intersection (CI) where the system can de-excite easily back to the ground state S0. Beyond 85°, we thus follow the further rotation in the ground state S0 until the metastable state 1-M (near 154°). Since steric hindrance sets in, we now allow the rotator (and stator) to be non-rigid in the ground state S0. This requires optimizing the structure at each fixed torsion angle (specified as a particular dihedral angle involving the C=C bond), which we did every 1° or 5° for Figure 3, but it could be done at larger intervals to save computational time. We observe again a positive total torque, approaching zero near the equilibrium state 1-M, as it should: so the rotation continues in the counterclockwise direction (as also confirmed in the more complete PES analysis [16]). The sudden switch to optimized geometries causes the observed discontinuities near 85° in Figure 3, which are therefore not physical.

To examine the C=C axis tilting impact on the torque, we project the torque of the rotator onto the Z direction ($\vec{T}_\parallel$). As shown in Figure S1 (a), the $\vec{T}_\parallel$ is almost the same to the torque projected onto the C=C bond, reflecting the axis tilting effect is minor. Figure S1 (b) shows the tilting angle of C=C axis from the Z direction to be unchanged at ~1.5° before 85° of the torsion angle due to the rigid rotation and to vary in the range of 0-21° in the relaxed geometries after 86° of the torsion angle. Therefore, the C=C axis tilting results in negligible impact on the torque we obtain.

In order to identify the atoms that dominate the unidirectional rotation, namely those atoms subject to the largest torques, we draw the profiles of the individual atomic torques of the rotor portion in Figure 4 as a function of the dihedral angle $\theta$. During the rotation in the S1 state from 15° to 86°, the C1 and C7 atom bear the largest torque. In the S0 state from 87° to 154°, the C1 atom basically bears the entire torque, all other atoms experiencing negligible torque. The C7 and C1 atoms belong to the same five-membered ring, and are connected together by C2 of C=C bonds, as shown in Figure 1. This shows that the optical response is still the cis-trans isomerization of the vinyl bond, but is blended into the structure of the more complex motor 1. After illumination, the C7 atom mainly undergoes most of the torque that drives the rotor portion to rotate in a counterclockwise direction until to the intermediate configuration at 85°, where the rotor and stator portions of the molecule are almost perpendicular to each other. Then after de-excitation from S1 to S0, the remaining rotation to 1-M will be driven by the C1 atom. As demonstrated above, the atomic torque analysis reveals very clearly that a single atom can dominate the rotation direction and can further deepen the study of the working mechanism of the motor. At the same time, in our example, the excitation also produces in the S1 state a distribution of atomic torques on multiple atoms with both negative and positive signs. This is unfavorable for starting a unidirectional rotation as these atoms will tend to rotate in opposite directions. Thus, our atomic torque profile can reveal the absolute preference of rotation direction so as to help to judge if the unidirectionality in a molecular motor is favorable or not.

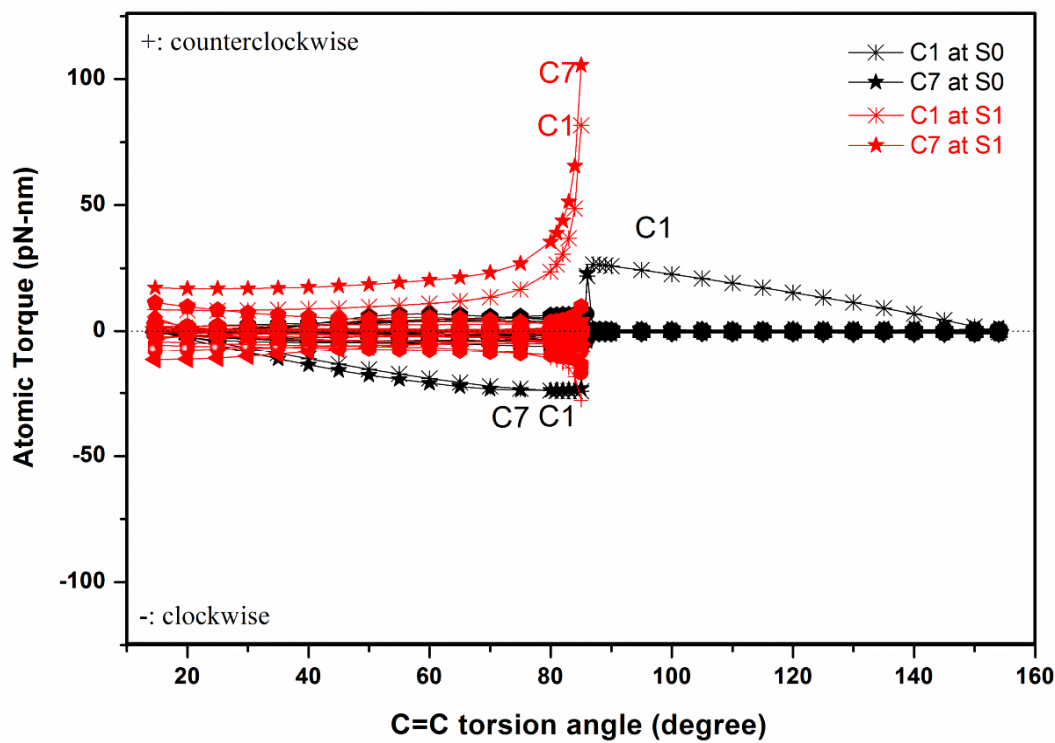

Figure 4. The atomic torques in the rotor portion in the S0 and S1 states as a function of the C=C double bond twisting from 1-P to 1-M. Only atom C1 and C7 are labelled, subject to a non-negligible torque.

**Conclusion**

In conclusion, we have developed an approach to use torque from DFT and TD-DFT calculations to predict the rotational behavior of a molecular rotor. The approach is successfully used in studying the photoisomerization of the light-powered molecular rotary motor 9-(2,4,7-trimethyl-2,3-dihydro-1H-inden-1-ylidene)-9H-fluorene twisting around its central C=C axis. Through vector summation of the torque acting on each atom and projected onto the rotational axis, we can optionally inspect the impact of a certain part within a motor on the unidirectional rotation with explicit rotation direction and magnitude. The unidirectional rotation can be realized by switching the S0 and S1 states around the conical intersection between them. The significant atomic torques are localized on two C atoms attached directly to the C=C axis in the rotor portion. Atomic torques dispersed onto multiple atoms after the illumination are not favorable for driving a unidirectional rotation.

Our work shows that the torque profile reveals the rotation direction, even though it is only an alternative expression of the ground and excited-state PESs. As potential energy is a scalar value, it cannot be decomposed and projected, unlike the calculated torque projected onto the rotational axis can clearly distinguish between the rotational directions in any configuration. The method thus can help experiments to improve motors or design motors with a better performance. The approach is very helpful in our on-going research to study the mechanical motions of other molecular motors.

**Acknowledgements**

The work described in this paper was supported by grants from the Research Grants Council of the Hong Kong SAR (Project Nos. CityU 11304415 and HKBU 12301814). We acknowledge the High Performance Cluster Computing Centre at the Hong Kong Baptist University, which receives funding from the Research Grants Council, the University Grants Committee of the HKSAR and HKBU, the National Supercomputing Center in Shenzhen for providing the computational resources, and Jing Xia for her participation in the early period of this work.

**References**

[1] A. J. Stone. "The Description of Bimolecular Potentials, Forces and Torques: the S and V Function Expansions", *Molec. Phys.* **36**, 241 (1978).

[2] S. L. Price, A. J. Stone, and M. Alderton. "Explicit Formulae for the Electrostatic Energy, Forces and Torques between a Pair of Molecules of Arbitrary Symmetry", *Molec. Phys.* **52**, 987 (1984).

[3] Book: "Theoretical Models of Chemical Bonding Part 4: Theoretical Treatment of Large Molecules and Their Interactions", ed. by Z.B. Maksic, Springer (1991).

[4] S. Mukherjee and A. Warshel, "Dissecting the Role of the γ-Subunit in the Rotary–Chemical Coupling and Torque Generation of F1-ATPase", *Proc Natl Acad Sci USA* **112**, 2746 (2015).

[5] S. Mukherjee, R. P. Bora, and A. Warshel, "Torque, Chemistry and Efficiency in Molecular Motors: a Study of the Rotary-Chemical Coupling in F1-ATPase", *Quarterly Rev. Biophys.* **48**, 395 (2015).

[6] K. Mislow, "Molecular Machinery in Organic Chemistry", *Chemtracts-Org. Chem.* **2**, 151 (1989).

[7] V. Balzani, M. Gömez-Löpez, and J. F. Stoddart, "Molecular Machines", *Acc. Chem. Res.* **31**, 405 (1998).

[8] J.-P. Sauvage, "Transition Metal-Containing Rotaxanes and Catenanes in Motion: Toward Molecular Machines and Motors", *Acc. Chem. Res.* **31**, 611 (1998).

[9] V. Balzani, A. Credi, F. M. Raymo, and J. F. Stoddart, "Artificial Molecular Machines", *Angew. Chem. Int. Ed.* **39**, 3348 (2000).

[10] B. L. Feringa, N. Koumura, R. W. J. Zijlstra, R. A. van Delden, N. Harada, "Light-driven monodirectional molecular rotor". Nature **401 (6749)**, 152–155(1999).

[11] J. Michl and E. C. H. Sykes, "Molecular Rotors and Motors: Recent Advances and Future Challenges", *ACS Nano* **3**, 1042 (2009).

[12] Torque in spacecraft: reaction wheel https://en.wikipedia.org/wiki/Reaction_wheel & attitude control https://en.wikipedia.org/wiki/Attitude_control

[13] Book: "Spacecraft Dynamics and Control: An Introduction" By Anton H. de Ruiter, Christopher Damaren, James R. Forbes, Wiley (2013).

[14] M. M. Pollard, A. Meetsma, and B. L. Feringa, "A Redesign of Light-Driven Rotary Molecular Motors", *Org. Biomol. Chem.* **6**, 507 (2008).

[15] A. Kazaryan, J. C. M. Kistemaker, L. V. Schäfer, W. R. Browne, B. L. Feringa, and M. Filatov, "Understanding the Dynamics Behind the Photoisomerization of a Light-Driven Fluorene Molecular Rotary Motor", *J. Phys. Chem. A* **114**, 5058 (2010).

[16] A. Kazaryan, Z. Lan, L. V. Schäfer, W. Thiel, and M. Filatov, "Surface Hopping Excited-State Dynamics Study of the Photoisomerization of a Light-Driven Fluorene Molecular Rotary Motor", *J. Chem. Theor. Comput.* **7**, 2189 (2011).

[17] T. Yanai, D. P. Tew, and N. C. Handy, "A New Hybrid Exchange–Correlation Functional using the Coulomb-Attenuating Method (CAM-B3LYP)", *Chem. Phys. Lett.* **393**, 51 (2004).

[18] I. V. Rostov, R. D. Amos, R. Kobayashi, G. Scalmani, and M. J. Frisch, "Studies of the Ground and Excited-State Surfaces of the Retinal Chromophore using CAM-B3LYP", *J. Phys. Chem. B* **114,** 5547 (2010).

[19] I. V. Rostov, R. Kobayashi, and R. D. Amos, "Comparing Long-Range Corrected Functionals in the Cis–Trans Isomerisation of the Retinal Chromophore", *Mol. Phys.* **110,** 2329 (2012).

[20] Gaussian 09, Revision C.01, M. J. Frisch, G. W. Trucks, H. B. Schlegel, G. E. Scuseria, M. A. Robb, J. R. Cheeseman, G. Scalmani, V. Barone, B. Mennucci, G. A. Petersson, H. Nakatsuji, M. Caricato, X. Li, H. P. Hratchian, A. F. Izmaylov, J. Bloino, G. Zheng, J. L. Sonnenberg, M. Hada, M. Ehara, K. Toyota, R. Fukuda, J. Hasegawa, M. Ishida, T. Nakajima, Y. Honda, O. Kitao, H. Nakai, T. Vreven, J. A. Montgomery, Jr., J. E. Peralta, F. Ogliaro, M. Bearpark, J. J. Heyd, E. Brothers, K. N. Kudin, V. N. Staroverov, R. Kobayashi, J. Normand, K. Raghavachari, A. Rendell, J. C. Burant, S. S. Iyengar, J. Tomasi, M. Cossi, N. Rega, J. M. Millam, M. Klene, J. E. Knox, J. B. Cross, V. Bakken, C. Adamo, J. Jaramillo, R. Gomperts, R. E. Stratmann, O. Yazyev, A. J. Austin, R. Cammi, C. Pomelli, J. W. Ochterski, R. L. Martin, K. Morokuma, V. G. Zakrzewski, G. A. Voth, P. Salvador, J. J. Dannenberg, S. Dapprich, A. D. Daniels, Ö. Farkas, J. B. Foresman, J. V. Ortiz, J. Cioslowski, and D. J. Fox, Gaussian, Inc., Wallingford CT (2009).

[21] M. Filatov and M. Olivucci, "Designing Conical Intersections for Light-Driven Single Molecule Rotary Motors: From Precessional to Axial Motion", *J. Org. Chem.* **79**, 3587 (2014).